\documentclass[
 reprint,
 superscriptaddress,
 amsmath,
 amssymb,
 aps,
 nofootinbib
]{revtex4-2}

\usepackage[utf8]{inputenc}
\usepackage[T1]{fontenc}
\usepackage{graphicx}
\usepackage{dcolumn}
\usepackage{bm}
\usepackage{natbib}
\DeclareUnicodeCharacter{2299}{$_\odot$}
\usepackage[normalem]{ulem}

\usepackage[usenames]{color}

\usepackage{hyperref}
\hypersetup{colorlinks,linkcolor={black},citecolor={blue},urlcolor={blue}}

\begin{document}

\title{Timing coincidence search for supernova neutrinos with optical transient surveys}

\author{Sean Heston}
\email{seanh125@vt.edu}
\affiliation{Center for Neutrino Physics, Department of Physics, Virginia Tech, Blacksburg, Virginia 24061, USA.}

\author{Emily Kehoe}
\affiliation{Department of Physics, Clarkson University, Potsdam, New York 13699, USA}

\author{Yudai Suwa}
\affiliation{Department of Earth Science and Astronomy, Graduate School of Arts and Sciences, The University of Tokyo, Tokyo 153-8902, Japan}
\affiliation{Center for Gravitational Physics and Quantum Information, Yukawa Institute for Theoretical Physics, Kyoto University, Kyoto
606-8502, Japan}

\author{Shunsaku Horiuchi}
\affiliation{Center for Neutrino Physics, Department of Physics, Virginia Tech, Blacksburg, Virginia 24061, USA.}
\affiliation{Kavli IPMU (WPI), UTIAS, The University of Tokyo, Kashiwa, Chiba 277-8583, Japan}

\date{\today}

\begin{abstract}
{Neutrinos allow the probing of stellar interiors during core collapse, helping to understand the different stages and processes in the collapse. To date, supernova neutrinos have only been detected from a single event, SN1987A. Most studies from then on have focused on two distance extremes; Galactic/local supernovae and all past cosmic supernovae forming the diffuse supernova neutrino background. We focus on the intermediate distance regime as a target for detecting core-collapse supernova neutrinos at next generation detectors like Hyper-Kamiokande. To quantify the significance of neutrino detections, we draw on expected discoveries by surveys of near galaxies as well as large synoptic surveys to monitor for optical counterparts of core-collapse supernovae. We find that detection prospects require approximately ten years of operation. We discuss how the ability of electromagnetic surveys to pinpoint the time of core collapse to within the timescale of hours is key for confident neutrino detections. Transient surveys like DLT40 which frequently observe nearby galaxies can help with such crucial information.}
\end{abstract}

\maketitle
\section{Introduction}\label{intro}

Neutrinos have been known to be vitally important for supernova explosions for over 50 years \cite{Arnett1966, ColgateWhite1966}. Current models of core-collapse supernovae (CCSNe) rely on neutrinos to revive the stalled bounce shock by heating matter behind the shock, allowing shock revival to occur \cite{Wilson1982, BetheWilson1985, bethe, burrows1993theory}. This is the delayed neutrino heating mechanism, and is the leading model explored in core-collapse simulations \cite{Takiwaki_2014, Lentz_2015, Melson_2015, Roberts_2016, Vartanyan2019Fornax, Powell_2020, Sandoval_2021, Vartanyan_2021}. Detecting neutrinos from CCSNe would allow the testing of current theories of core-collapse processes, probing of neutrino properties at energies and densities not achievable on Earth, and could tell us about the evolution of the protoneutron star (PNS) formed during core collapse \cite{Scholberg_2004_review, Kotake_2006, Lund_2010, janka2012, Scholberg_2012, Cherry_2013, Kistler_2013, Tamborra_2013, Tamborra_2014, Tamborra_2014b, Mirizzi_2016, Kneller_2017, Horiuchi_2017, Wright_2017, Horiuchi_2018, Mori_2021,  Chang_2022, Nakazato_2022}. However, only around two dozen supernova neutrinos have been detected to date, all from SN1987A \cite{1987A1, 1987A2, 1987A3}. With only a small repository of events, many more detections are needed for a better understanding of CCSNe processes and physics. 

The two major areas of research on detecting CCSNe neutrinos focus on radically different distance regimes, the galactic/local regime and the diffuse supernova neutrino background (DSNB) \cite{Ando_2004_DSNB, Vitagliano_2020_DSNB}. In the former, the target are CCSNe occurring within the Milky Way and its satellites, with most studies looking at CCSNe near the Galactic Center ($d\,{\sim}\,10$~kpc). At these distances, next generation neutrino detectors such as Hyper-Kamiokande (HK) \cite{hyperkamiokande} will detect $\mathcal{O}(10^5)$ neutrino events \cite{hyperkamiokande, Nagakura_2020, Abe_2021}. On the other hand, the DSNB regime looks at the neutrino flux from all past cosmic CCSNe. The DSNB flux is predicted to be detectable, e.g., at Super-Kamiokande (SK) and HK \cite{hyperkamiokande, Horiuchi_2017_DSNB, deGouvea_2020, Libanov_2022}. However, both of these two regimes have unique challenges. For galactic CCSNe, the rate at which they occur is very slow on human timescales. Estimates put the galactic CCSNe rate at around $1{-}3$ per century \cite{Diehl_2006, Adams_2013}, and the only remedy is patience. The DSNB's challenge is that being an isotropic signal constant in time, it does not correlate with specific objects. Also, it must compete against various backgrounds \cite{SKbkgd,SKIV_DSNB}, although this has been largely mitigated with the recent gadolinium upgrade at SK \cite{GdDoping, SuperKGdLoading}. As an additional third approach, we focus on a relatively unexplored distance regime that lies between local CCSNe and all past cosmic CCSNe. 

Due to transient surveys, CCSNe are now routinely discovered in the intermediate distance regime. However, most of these CCSNe still occur at large distances such that their resulting detectable neutrino fluxes are much less than one event at HK. In order to account for this low neutrino flux, we explore longer search time periods during which rarer but closer CCSNe will occur. We estimate the ``combined detection rate'' which represents how many neutrino events we can expect over an extended observation period. For closer CCSNe, targeted surveys like DLT40 (explained in Sec.~2 of Ref.~\cite{Tartaglia_2018}) which preferentially targets nearby galaxies with high cadence will likely provide early observations of the CCSN light curve. For further CCSNe ($D>40$ Mpc) large-sky surveys such as the upcoming Legacy Survey of Space and Time (LSST) out of the Vera C.~Rubin Observatory \cite{LSST} will provide coverage and cadence for optical detections. Such coincident optical detections will constrain the time of core collapse and reduce the time window to search for neutrinos, thus helping to improve neutrino detection significance. However, the number of CCSNe that we can study while avoiding an ``always on'' mode (i.e., the DSNB regime) is limited by how small the uncertainty is for the estimates of time of core collapse. For example, assuming the uncertainty to be one day allows us to study up to some 365 CCSNe per year, while more CCSNe can be considered if the uncertainty in time of core collapse is reduced. 

The primary goal of this paper is to quantify the search for CCSN neutrinos in the intermediate distance regime, namely, how long of a search period is necessary to achieve neutrino detections and to identify whether there is ${>}1$ expected neutrino detection per year coincident with CCSNe without going into the continuous DSNB regime. For this purpose we assume a future 2-tank HK neutrino detector with and without gadolinium and a range of uncertainties in the estimated time of core collapse ($\Delta t$) from $\Delta t\;{=}\;1$ day to $\Delta t\;{=}\;1$ hour. We find that there is a region of interest in which there is ${>}1$ CCSNe neutrino event for uncertainties in time of core collapse ranging from 1 to ${\sim}$3 hours. However, the number of background events during the same time period is larger than the expected signal. Therefore, we suggest that in order to get a good signal-to-noise ratio (SNR), data must be collected over more than one year such that only the closest observed CCSNe can be considered. For example, we find that $8-10$ years yield  significant improvements in detection prospects. 

The intermediate distance regime has previously been explored in Refs.~\cite{Ando_2005, Kistler_2011, Nakamura_2016}. Our study goes beyond these studies in various ways. References \cite{Ando_2005, Kistler_2011} focus on using megaton scale detectors to achieve a detection horizon for CCSNe with neutrinos alone of several Mpcs. Optical coincidence detection is mentioned in Ref.~\cite{Ando_2005} and they reach a similar conclusion in that the uncertainty in time of core collapse needs to be reduced below one day. Reference \cite{Nakamura_2016} considers a larger HK detector (${\sim}0.5$ Mton), with and without gadolinium doping. They consider ``minibursts'' of a few events within a 10 second time window, giving a horizon distance of a few Mpc. We build on these studies with detailed modeling of the rates of CCSN discoveries and consider smaller neutrino detector configurations, allowing us to estimate more up-to-date single neutrino detections with optical timing coincidence in upcoming experimental setups. 

The paper is laid out as follows: Sec.~\ref{bkgd} overviews the theory and detection of the neutrino and optical signals for CCSNe, Sec.~\ref{Modeling} covers the modeling of observable distance distributions of CCSNe for the timing coincidence search as well as a comparison between theory and observations for a currently running transient survey, and Sec.~\ref{Results} discusses the probability of detecting a single neutrino from a CCSNe, the significance of detection for a single neutrino, defines the single-year search area, as well as finds the theoretical neutrino event yields for many years of observations. We summarize and conclude in Sec.~\ref{summary discussion}. 

\section{Supernova Modeling and Observations}\label{bkgd}

We first cover core-collapse theory and our analytic model of neutrino emission from core collapse. Next, we go over detector parameters for neutrino signals at HK. Finally, we go over the parameters of the optical transient surveys we consider.

\subsection{Core-collapse neutrinos}\label{Core-Collapse Theory}

Core collapse is a process that occurs for stars with initial masses greater than ${\sim}8$  $\mathrm{ M_\odot}$. Core collapse is brought about by gravity overcoming the electron degeneracy pressure that supports the stellar core. Once overcome, the core undergoes free-fall collapse until nuclear densities are reached, and the core stiffens. This stiffening causes a bounce shock from matter rebounding off the PNS formed by the collapsed core which, if it reaches the stellar surface, will cause a supernova explosion \cite{janka2012b}. However, simulations have shown that this bounce shock will not reach the surface due to energy loss, and it will become a stalled accretion shock, eventually leading to black hole formation for the PNS unless the shock is revived \cite{Burbridge1957}. Later simulations showed that if neutrino heating was introduced, it could provide the necessary energy for a successful explosion \cite{ColgateWhite1966}, as neutrinos are produced in copious amounts during the stages of core collapse \cite{RevModPhys.75.819, Sato_1987} and they only need to impart ${\sim}1\%$ of their total energy into the shock for a successful explosion. The current model of neutrino heating in CCSNe is a delayed heating mechanism \cite{Wilson1982, BetheWilson1985, bethe}. In this model, the bounce shock propagates outward until it stalls due to energy loss. Neutrinos then interact with matter behind the shock, heating it and causing outward pressure. This outward pressure can eventually revive the shock under certain conditions, leading to a successful SN explosion.

We assume an analytic model of the time-integrated neutrino emission for our modeling. This model is applied to all CCSNe that we consider. Theoretical models of CCSNe suggest a pinched Fermi-Dirac spectrum for neutrino emission. For our spectrum, we use the normalized spectrum of Ref.~\cite{PinchedSpectrum}, 
\begin{equation}\label{NeutrinoSpec}
    F(E_{\bar{\nu}_e})=\frac{L_{\bar{\nu}_e, \mathrm{tot}} E_{\bar{\nu}_e}^\alpha}{\langle E_{\bar{\nu}_e} \rangle ^{2+\alpha}} \frac{(\alpha+1)^{(\alpha+1)}}{\Gamma(\alpha+1)} \, \mathrm{Exp}\left[-(\alpha+1)\frac{E_{\bar{\nu}_e}}{\langle E_{\bar{\nu}_e} \rangle} \right],
\end{equation}
\noindent where $L_{\bar{\nu}_e, \mathrm{tot}}$ is the total electron anti-neutrino luminosity, $\langle E_{\bar{\nu}_e} \rangle$ is the average neutrino energy, $\alpha$ is a parameter that allows for pinched, antipinched, or Fermi-Dirac distributions, and $\Gamma$ is the gamma function. This equation comes from Monte Carlo studies of CCSNe neutrino emission \cite{PinchedSpectrum}. The parameters of the spectrum vary over the emission time due to core-collapse stages , progenitor dependence, and asymmetric processes. For simplicity, we assume constant values for these parameters that are motivated by CCSNe simulations \cite{PinchedSpectrum, Totani_1998, Thompson_2003, Sumiyoshi_2005, Nakazato_2013, Suwa_2019, Bollig_2021}. These representative values are: $L_{\bar{\nu}_e, \mathrm{tot}}\,{\sim}\,5\times10^{52}   \mathrm{erg}$,  ${\langle E_{\bar{\nu}_e} \rangle} \,{\sim}\, 15   \mathrm{MeV}$, and $\alpha\,{\sim}\,2.3$. References \cite{PinchedSpectrum, Totani_1998, Thompson_2003, Sumiyoshi_2005, Nakazato_2013, Suwa_2019, Bollig_2021} suggest a higher value of $\alpha$, however the resulting change to the event rate from this choice is small. We also assume that neutrinos are massless and thus no flavor oscillation occurs. This is justified for our time-integrated emission since the flavor dependence becomes significantly smaller a few seconds after the supernova explosion (e.g., Ref.~\cite{Suwa_2019}). 

\subsection{Neutrino detection at HK}\label{Neutrino Detection}

For neutrino events, we model detection at a one-tank and a two-tank HK with a total fiducial volumes of 187 kton and 375 kton, respectively. We also model detector configuration with and without Gd doping, allowing for efficient neutron tagging of events \cite{GdDoping}. Currently, excavation is ongoing for a single tank in Japan, but there are prospects of another tank in a different country such as South Korea \cite{hyperkamiokande}. Similarly Gd doping is not part of the default HK design, but may be doped in the future just as SK was. 

As we assume that the two-tank HK will be Gd doped, we have to take into account the neutron tagging efficiency it brings. Without Gd doping, thermal neutrons from inverse beta decay (IBD) mostly capture onto free protons, but the timescale for this compared to the prompt positron emission is quite long. The energy of the $\gamma$ released from the neutron capturing onto a free proton is also not that large compared to detection capabilities. However, with Gd dissolved into the water, the neutrons will capture onto Gd as it has a much larger cross section for thermal neutron capture and happens much faster, ${\sim} 30 \,  \mathrm{{\mu s}}$. The Gd then deexcites releasing an 8 MeV $\gamma$ cascade which is detectable \cite{GdDoping}. In this work, we take the efficiency for neutron tagging from Gd doping to be 90\%. This reduces the background by a factor of around 3.

We model events only through the IBD channel, which is the dominant reaction channel for CCSNe neutrinos in HK. We assume a detection energy window of $11{-}30$ MeV ($16{-}30$ MeV) with (without) Gd doping. These bounds come from avoiding backgrounds such as reactor neutrinos (${<}10$ MeV) and atmospheric neutrinos and muons (at high energies). The cross section of IBD is taken from Refs.~\cite{Vogel_1999, Strumia_2003, Ricciardi_2022}, given by 
\begin{equation}\label{CrossSec}
    \sigma(E_\nu)=9.5\times10^{-44} \left( 1-6 \, \frac{E_{\bar{\nu}_e}}{M} \right) \left[ \frac{E_{\bar{\nu}_e}-\Delta}{\mathrm{MeV}} \right]^2 \mathrm{cm^2},
\end{equation}
\noindent where, $E_{\bar{\nu}_e}$ is the electron anti-neutrino energy, $M$ is the average nucleon mass ($M= (m_n+m_p)/2$), and $\Delta$ is the nucleon mass difference. 

The target of IBD in HK is the free protons in the water. This gives the target number simply being the number of hydrogen atoms in the water. Using Eqs.~(\ref{NeutrinoSpec}) and (\ref{CrossSec}), we can calculate the neutrino event yield via 
\begin{equation}\label{NeutrinoYield}
    N_{\bar{\nu}_e} = \frac{N_{\rm t}}{4\pi D^2}\int^{E_{\rm high}}_{E_{\rm low}} \sigma (E_{\bar{\nu}_e}) \, F(E_{\bar{\nu}_e}) \,dE_{\bar{\nu}_e},
\end{equation}
\noindent where $N_{\rm t}$ is the number of target protons, $D$ is the distance to the CCSNe, and $E_{\rm low}$ and $E_{\rm high}$ are the bounds of the energy range of interest. Gd doping allows for neutron tagging with up to ${\sim}90\%$ efficiency, reducing backgrounds and increasing energy range. 

\subsection{Optical surveys: DLT40, ASAS-SN, LSST}

As we will be looking for a timing coincidence between observed CCSNe and neutrino signals, we need a collection of observed CCSNe. For our purposes, we look at DLT40 (described in Ref.~\cite{Tartaglia_2018}). DLT40 is a targeted survey, i.e., targets a precompiled list of galaxies, which tries to discover as many SNe within a day of explosion occurring within approximately 40 Mpc. Thus, the target includes Milky Way globular clusters, galaxies of all manners within $D\sim11$ Mpc, and high star-forming galaxies within $D\sim40$ Mpc. DLT40 uses PROMPT 0.4 m telescopes \cite{PROMPT} with a limiting magnitude of ${\sim}19$ in the $r$-band. There is one telescope at the Cerro Tololo Inter-American Observatory in Chile, a second in Australia, and a third in Canada. 

For larger sky surveys, we consider two surveys, one ongoing and one future. The first is the All-Sky Automated Survey for Supernovae (ASAS-SN). ASAS-SN has telescopes all around the world, giving it full sky coverage. They mainly observe SNe in the $g$-band with a limiting magnitude of 18 \cite{ASAS-SN2022} and an average cadence of $2{-}3$ days \cite{ASAS-SN_Photometry}. For the next generation survey, we consider LSST. LSST is expected to have a sky coverage of around 20,000\;$\mathrm{deg}^2$ (slightly less than half of the sky). LSST is planning to collect SNe observations in the $r$-band with a single-visit limiting magnitude of $\sim$24.5 (we assume a limiting magnitude of 24 which should only decrease distant CCSNe observations). LSST has a planned revisit time of three days on average per $10,000 \; \mathrm{deg}^2$ with two visits per night \cite{LSST}. The $10,000 \; \mathrm{deg}^2$ is assumed to be the area visible at any given time of year.

\section{Modeling CCSN discovery}\label{Modeling}

In this section, we work out the theoretical modeling of the distance distribution of CCSNe that ASAS-SN and LSST can observe. We do not do modeling for DLT40 as that is a targeted survey. In order to accomplish this modeling, we follow the methods of Ref.~\cite{Lien&Fields2009}, with changes to some of the input parameters. For all calculations, we use a $\mathrm{\Lambda}$CDM model with $\Omega_\Lambda=0.7$, $\Omega_\mathrm{m}=0.3$, and $H_0=70 \; \mathrm{km \, s^{-1} \,Mpc^{-1}}$. 

We start with using luminosity functions of each CCSNe subtype along with each survey's observational parameters to calculate the detection efficiency of each survey. Then, using a cosmic comoving CCSNe rate (CSNR), $\mathcal{R}_{\mathrm{SN}}(z)$, that is proportional to the cosmic star formation rate (CSFR) \cite{Madau_1998}, we estimate the idealized CCSNe detection rate of each survey, given by \cite{Lien&Fields2009}
\begin{equation}\label{Ideal Rate Eqn}
        \frac{dN_{\mathrm{SN}}}{d\Omega dt_{\mathrm{obs}} dz} = \mathcal{R}_{\mathrm{SN}}(z)\frac{r^2_{\mathrm{com}}}{1+z}\frac{dr_{\mathrm{com}}}{dz},
\end{equation}
\noindent where $r_\mathrm{com}$ is the comoving distance. Taking into account observational effects, covered in Sec.~\ref{Detection Efficiency}, we retrieve the estimated observed CCSNe detection rate per solid angle per redshift bin. Using this detection rate along with each survey's sky coverage and survey time, we obtain the theoretical distance distribution of observed CCSNe for each survey for an arbitrary choice of redshift bin size.

\subsection{Luminosity functions} \label{Lum funcs}

We update the luminosity functions adopted in Ref.~\cite{Lien&Fields2009}, which is based on Ref.~\cite{Richardson_2002} from 2002, with the latest luminosity functions from the same group updated in 2014 \cite{Richardson_2014}. Both Refs.~\cite{Richardson_2002,Richardson_2014} use the same source catalog (the Asiago Supernova Catalog \cite{AsiagoCatalogue}), however the number of SNe in the catalog increased more than threefold between 2002 to 2014. Due to the larger amount of data, Ref.~\cite{Richardson_2014} should more accurately describe the luminosity distributions of CCSNe, especially the tails containing rarer explosions. 

The luminosity functions are given in terms of absolute magnitude, which we convert into apparent magnitude via $M_x(z)=m_x-\mu(z)-K_x(z)-\eta_{xy}$ for survey passband $x$ and luminosity functions measured in passband $y$. Here, $M_x$ is the absolute magnitude, $m_x$ is the apparent magnitude, $\mu(z)$ is the standard distance modulus, $K_x(z)$ is the $K$-correction which takes redshift effects into account, and $\eta_{xy}$ is a color correction due to the fact that the luminosity functions are observed in a different passband than the observational passbands of ASAS-SN and LSST. The $K$-correction is found using Eq.~(C1) of Ref.~\cite{Lien&Fields2009}. This calculation requires the spectral shapes of the supernovae and the passband sensitivities of the bands of interest. The spectral shapes are taken to be thermal blackbody spectra as prescribed by Ref.~\cite{Dahlen&Fransson}. The passband sensitivities for the \textit{B-,g-,r-}bands are taken from Ref.~\cite{StandardPhotometric}. The color correction is given by Eq.~(3.8) of Ref.~\cite{Lien&Fields2009}. This also requires the spectral shapes and the passband sensitivities, as well as a zero-point correction due to the fact that different passbands have different zero points. The zero-point corrections are calculated using the star $\mathrm{BD}\,{+}\,17^{\circ}4708$ \cite{Landolt2007,Kessler2009}, which is the zero point of the SDSS magnitude system \cite{Fukugita_1996}.

\subsection{Detection efficiency}\label{Detection Efficiency}

\begin{figure}[t]
    \centering
    \includegraphics[width=8.6cm]{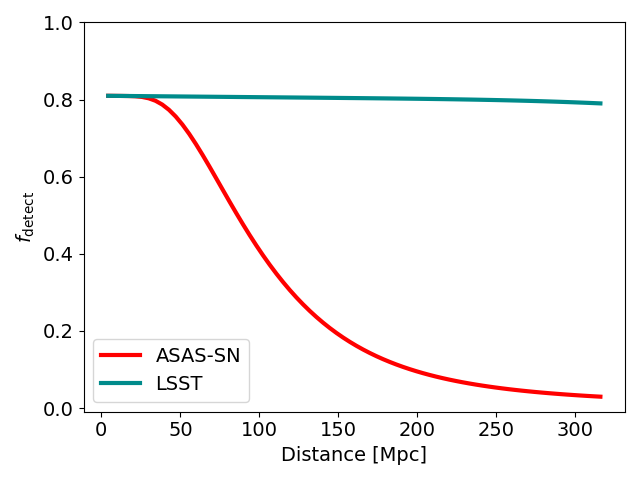}
    \caption{Idealized detection efficiency plot for ASAS-SN and LSST. The detection efficiency is estimated by the values of $f_{\rm detect}$ for each survey, which is based on the CCSN luminosity functions of Ref.~\cite{Richardson_2014}, CCSNe type weights of Ref.~\cite{Loss2}, the sensitivity limits of each survey, and a redshift-dependent dust scheme \cite{DustObscuration}.}
    \label{Survey Detection Efficiencies}
\end{figure}

With the luminosity functions now defined in terms of apparent magnitude in survey passbands, we can find the detection efficiency as a function of distance of each survey. The detection efficiency represents the observable fraction of supernovae that the surveys can detect depending on their limiting magnitude and dust extinction along the line of sight to the supernova. 

For dust extinction, we adopt what is suggested by Ref.~\cite{DustObscuration}. This is stronger than what is used in Ref.~\cite{Lien&Fields2009}, which follows Ref.~\cite{Mannucci_2007}, but is based on recent nearby supernovae observations making it a better model for our needs. We also need to account for the fact that there are different spectral subtypes of CCSNe, some of which occur more frequently than others. In order to correct for subtype distributions, we use volumetric weight fractions, which describe how many CCSNe in a volume-limited sample of supernovae observations correspond to each subtype. We use those from the Lick Observatory Supernova Search \cite{Loss2} which finds a subtype ratio Ibc:IIP:IIL:IIn of 0.27:0.57:0.08:0.07. Then, we define the detection efficiency to be \cite{Lien&Fields2009}:

\begin{equation}
    \label{fdetect}
    f_\mathrm{detect}(z;m^\mathrm{sn}_\mathrm{lim})=f_\mathrm{maglim}(z;m^\mathrm{sn}_\mathrm{lim})f_\mathrm{dust}(z),
\end{equation}

\noindent where $f_\mathrm{maglim}(z;m^\mathrm{sn}_\mathrm{lim})$ is the fraction of CCSNe detectable as a function of redshift due to the limiting magnitude of a survey and $f_\mathrm{dust}(z)$ is the fraction of CCSNe not obscured by dust extinction as a fraction of redshift \cite{DustObscuration}. The survey detection efficiencies are shown out to ${\sim}300$ Mpc in Fig.~\ref{Survey Detection Efficiencies}. Both ASAS-SN and LSST converge to an ideal detection efficiency ${\sim}0.8$ at low distances, because the model of dust extinction we use assumes ${\sim}20\%$ are not detected. As seen in the figure, LSST remains mostly complete out to over 300 Mpc, while ASAS-SN quickly becomes inefficient past 50 Mpc. This is due to the fact that ASAS-SN has a lower limiting magnitude of $g \sim 18$, while LSST has limiting magnitude of $r \sim 24$. 

\subsection{Observed differential CCSNe rate and distance distribution}\label{Observed Rate and Distance Distribution}

The next step for modeling the number of observable CCSNe is to derive the CSNR, $\mathcal{R}_\mathrm{SN}$. We estimate the CSNR to be proportional to the CSFR, $\dot{\rho}_*$ \cite{Madau_1998}. We use the linear piecewise fit of Ref.~\cite{Hopkins_2006}. The linear fit is used as it has a better match to observations at the small distances we are concerned about in this work. Assuming a Salpeter A initial mass function \cite{SalpeterA} and taking the initial stellar mass of CCSNe progenitors to be between $8{-}50$ $\mathrm{ M_\odot}$, we find that $\mathcal{R}_{\mathrm{SN}}/\dot{\rho}_*=0.00914 \;\mathrm{M_\odot^{-1}}$.

\begin{figure}[t]
    \includegraphics[width=8.6cm]{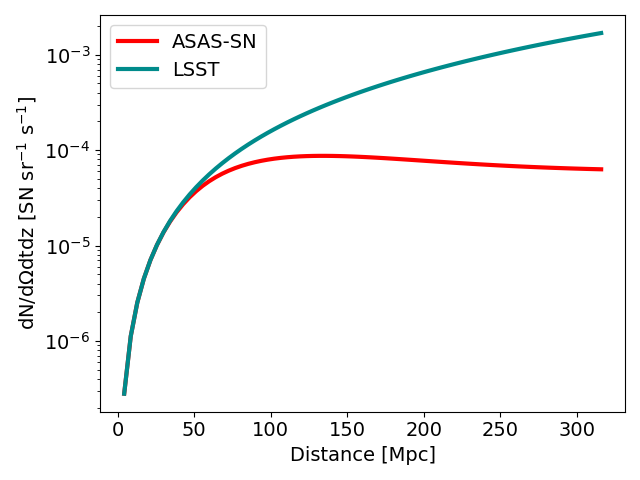}
    \caption{Idealized differential CCSN rate observed at ASAS-SN and LSST calculated with Eq.~(\ref{Observed Rate Eqn}) using the linear CSFR fit of Ref.~\cite{Hopkins_2006}.}
    \label{Linear SN Rates}
\end{figure}

The idealized observation rate of CCSNe is then just given by Eq.~(\ref{Ideal Rate Eqn}). However, we have to take observational effects into account that will decrease the observed rate compared to the ideal rate. These observational factors are what comprise $f_{\rm detect}$. Therefore, the observed CSNR per solid angle and redshift bin is then given by \cite{Lien&Fields2009}

\begin{equation}\label{Observed Rate Eqn}
\begin{split}
    \Gamma_\mathrm{SN,obs,x}(z) &\equiv \frac{dN_{\mathrm{SN,obs,x}}}{dt_{\mathrm{obs}}\,dz\,d\Omega} \\
    &=\mathcal{R}_\mathrm{SN}(z)\,f_\mathrm{detect,x}(z;m_\mathrm{lim}^\mathrm{sn})\,\frac{r(z)_{\rm com}^2}{1+z}\,\frac{dr_{\rm com}}{dz}.
\end{split}
\end{equation}

This observed CSNR is shown in Fig.~\ref{Linear SN Rates}. LSST follows a mostly volumetric increase out to ${\sim}$300 Mpc, whereas ASAS-SN begins to flatten past ${\sim}50$ Mpc. This is due to the fact that ASAS-SN is inefficient at these large distances, so even though there are more CCSNe occurring at these distances, the vast majority of them are not observable for ASAS-SN due to its limiting magnitude.

\begin{figure}[t]
    \includegraphics[width=8.6cm]{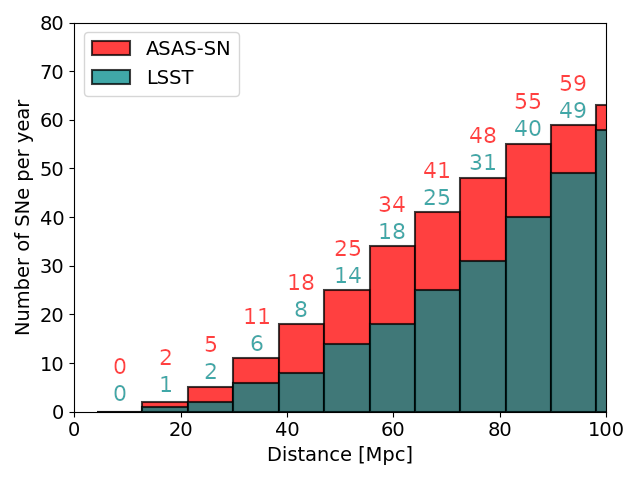}
    \caption{Modeled differential histogram over distance bins for the observable CCSNe per year for ASAS-SN and LSST within 100 Mpc. ASAS-SN has a larger sky coverage so it sees more CCSNe at these distances even though it has a smaller detection efficiency.}
    \label{100 Mpc Distance Distribution}
\end{figure}

Using Eq.~(\ref{Observed Rate Eqn}), we can work out the theoretical observed differential number of CCSNe per redshift bin per year over each survey's scan area. This is done by multiplying Eq.~(\ref{Observed Rate Eqn}) by the respective scan area of each survey $\Delta \Omega$, the observation time $\Delta t_{\rm obs}$ (one year), and the redshift bin size $\Delta z$ (=0.002). The fineness of the redshift binning was chosen to represent a velocity space uncertainty of ${\pm}300$ km/s. The CCSNe for each survey is calculated according to this method and most are shown in Fig.~\ref{100 Mpc Distance Distribution}. We see that ASAS-SN's closest CCSNe occur at smaller distances than the closest of LSST. This comes from the fact that ASAS-SN, even though it is more inefficient than LSST at large distances, has a larger sky coverage which causes it to see more CCSNe at the closest distances. 

\begin{figure}[t]
    \centering
    \includegraphics[width=8.6cm]{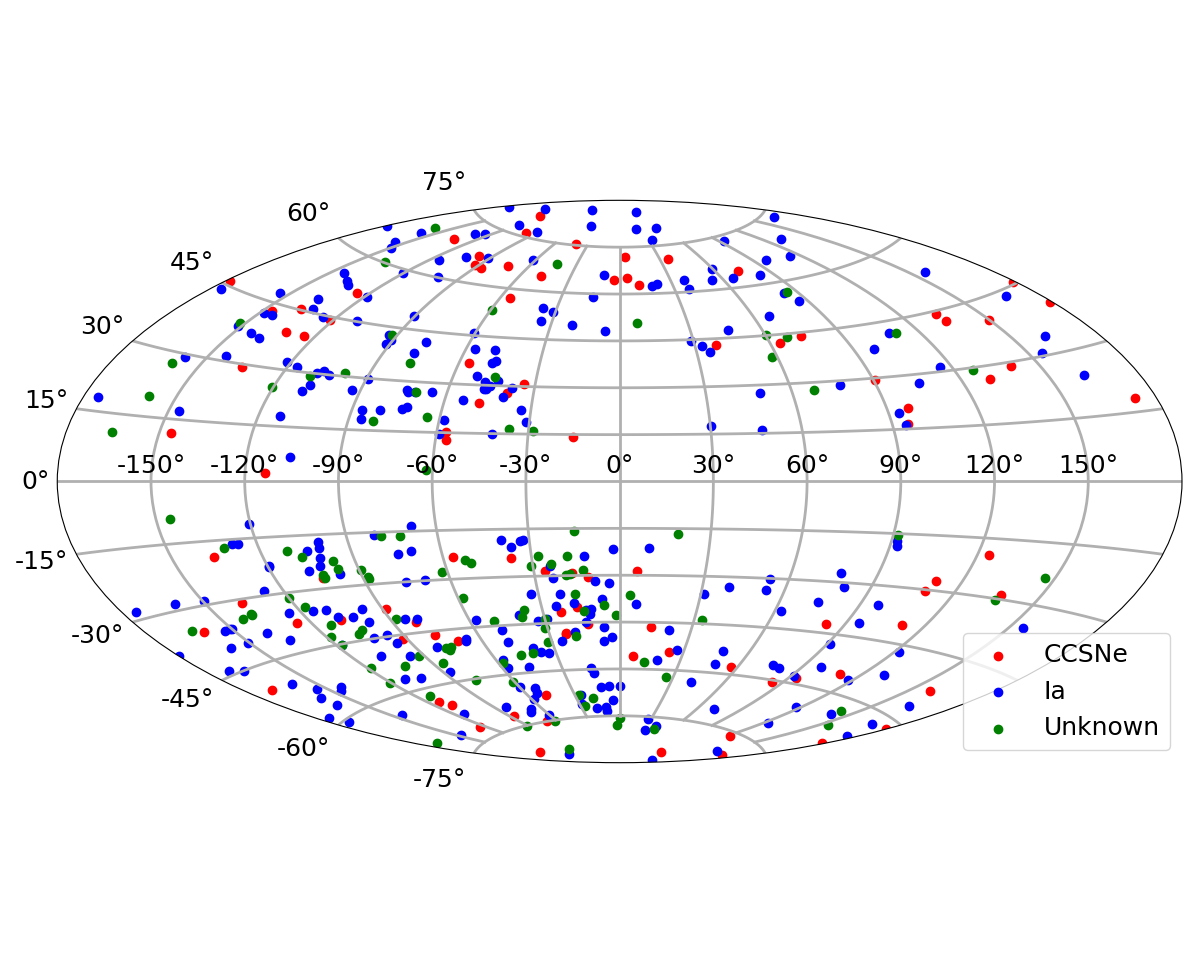}
        \caption{Scatter plot of the observed CCSNe from ASAS-SN for 2018 and 2019. Red denotes confirmed CCSNe, blue is Type Ia SNe, and green is unknown type SNe. The scatter plot is on a Aitoff projection in galactic coordinates.}
    \label{ASAS-SN SNe Coverage}
\end{figure}

Compared to ASAS-SN, our rate estimates are large. For example, during the years $2018{-}2019$ (which are more complete than the most recent $2020{-}2021$ \cite{ASAS-SN2022}), ASAS-SN observed a total of 495 SNe, of which 289, 96, and 110 are classified spectrally as Type Ia, CCSNe, and unknown, respectively.\footnote{\url{https://www.astronomy.ohio-state.edu/asassn/sn\_list.html}} Of the confirmed CCSNe, 2, 9, and 22 are within 20, 40, and 60 Mpc. By comparison, we estimate 2, 36, and 95, which are significantly higher. There are several reasonable reasons for the difference. First, the sky coverage of ASAS-SN is less than the full sky: in Fig.~\ref{ASAS-SN SNe Coverage} we show the sky positions of all observed SNe by ASAS-SN from 2018 and 2019, regardless of type and redshift, in Aitoff projection in galactic coordinates, which shows a clear dearth of SNe along the Milky Way plane. With a sky coverage reduction factor of ${\sim}3$ we can match ASAS-SN out to ${\sim}25$ Mpc. Nevertheless, we are surely likely to be missing additional effects of dust, host galaxy confusion, weather/seeing, runtime incompleteness, etc, which we do not attempt to correct but can all reduce the detection efficiency. 

\section{Results}\label{Results}

In this section, we explore the prospects of detecting neutrinos from intermediate distance CCSNe. We first explore the probability of detecting a neutrino event from a CCSN with different detector configurations. Next, we find the significance of detecting a single neutrino with a certain expected number of background events that is dependent on the search window size. Finally, we find the expected CCSNe neutrino event rates over differing years spent collecting observations. As detected events are integers and we are summing fractional events, the actual number of events may be lower or higher \cite{Suwa_2022}.

\subsection{Detecting a single neutrino event}

\begin{figure}[t]
    \centering
    \includegraphics[width=8.6cm]{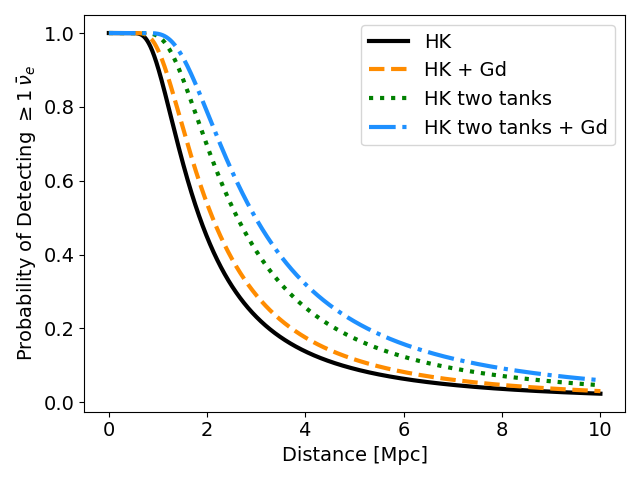}
        \caption{Probability of detecting at least one neutrino event from a CCSN as a function of distance using different detector configurations. Gd doping has a larger horizon as it allows for a wider energy band; $E_\mathrm{low}=11$ MeV instead of 16 MeV.}
    \label{Prob at least 1 event}
\end{figure}

We use Poisson statistics to find the probabilities and significances of detecting a single neutrino event from a CCSN with different detector setups. Specifically, what we care about is the cumulative distribution function (CDF), 
\begin{equation}\label{Poisson CDF}
    Q\left( \lfloor k+1\rfloor, \lambda \right)=\frac{\Gamma\left(\lfloor k+1 \rfloor, \lambda\right)}{\lfloor k \rfloor !},
\end{equation}
\noindent where $\Gamma$ is the upper incomplete gamma function, $\lfloor x \rfloor$ is the floor function of $x$, $k$ is the integer index corresponding to the number of events, and $\lambda$ is the expected number of events. The CDF itself represents the cumulative probability of detecting $k$ or less events. With the CDF, we can then find the probability of detecting at least one neutrino event as $1-Q\left( \lfloor 1\rfloor, \lambda \right)$, where we set $k=0$. For the values of $\lambda$, we calculate the expected number of events using Eq.~(\ref{NeutrinoYield}) for the following detector configurations: one tank HK without Gd, one tank HK with Gd, two tank HK without Gd, and two tank HK with Gd. The resulting probabilities are shown in Fig.~\ref{Prob at least 1 event}. This detection horizon does not take into account backgrounds and the differences arise from the different energy ranges for no doping ($E_{\mathrm{low}}=16$ MeV) versus Gd doping ($E_{\mathrm{low}}=11$ MeV) and detector mass. We keep $E_{\mathrm{high}}$ fixed to 30 MeV. 

All configurations should be able to detect at least one neutrino from a CCSN occurring within ${\sim}$1 Mpc. The only large galaxies within 1 Mpc are Andromeda and Triangulum. Estimating their core-collapse rates from their star formation rates and adding that to the Milky Way core-collapse rate, we retrieve that within 1 Mpc CCSNe should occur at a rate of $3{-}5$ per century. Using the star formation rates to estimate the core-collapse rate at these small distances gives us a lower limit. Galaxies at these small distances are targets for DLT40.

Next, we find the significance of detecting a single neutrino event. For this, we again use the CDF with $k=0$, but now $\lambda=R_{\mathrm{bkgd}}\Delta t$ where $R_{\mathrm{bkgd}}$ is the background rate and $\Delta t$ is the search window size. This means that the detection significance is just the probability of detecting zero neutrino events that come from backgrounds,

\begin{equation}\label{Eqn Detection Significance}
\begin{split}
    P(k=0, \lambda=R_\mathrm{bkgd}\Delta t)&=\frac{\Gamma\left(\lfloor 1 \rfloor, R_\mathrm{bkgd}\Delta t \right)}{\lfloor 0 \rfloor !}, \\
    & = e^{- R_\mathrm{bkgd} \Delta t}.
\end{split}
\end{equation}

For the background rates, we use the rates of Ref.~\cite{SKbkgd}, an offline DSNB search of SK data. Although HK will have a lower photomultiplier tube (PMT) coverage than SK, which can result in higher backgrounds, the use of higher efficiency PMTs in HK can partially offset this issue. For our purposes, we will estimate backgrounds based on SK runs, but further work will be necessary to quantify the impacts of the HK design. Our main focus is to explore to what extent backgrounds could be a major issue. We also need to consider the contributions to the background from the DSNB, as this is a CCSN neutrino signal that we may coincidentally detect in the time window for a detected CCSN. For a DSNB estimate, we use Ref.~\cite{Moller_2018}, which finds the DSNB flux at a two-tank HK with Gd doping, and ends up contributing 20 events per year. 

We use two different SK background rates: one is the background rate from the SK-II run and another is the average background rate of the first three SK runs. We expect HK's background rate to be most similar to SK-II, as that is the period when SK had lower PMT coverage. As a side note, the studies of DSNB neutrinos at SK (e.g., Refs.~\cite{SKbkgd, SKIV_DSNB}), only model backgrounds down to 15 MeV whereas we model detection down to 11 MeV. At these lower energies between $11{-}15$ MeV, spallation should rise and can increase the backgrounds even further. The largest background in the energy range of interest is atmospheric $\nu_\mu$ and $\bar{\nu}_\mu$ undergoing charged-current (CC) reactions inside the detector. Other backgrounds include atmospheric $\nu_e$ and $\bar{\nu}_e$ CC events, atmospheric $\mu$ and $\pi$, neutral current elastic scattering events from atmospheric $\nu$, and spallation. Averaging over SK-I through SK-III (just SK-II) and including 20 DSNB events \cite{Moller_2018}, we extract an average background rate of 0.00163 (0.00155) events per day per kton of detector material in the energy range of interest for a detector without Gd doping.

We also extrapolate these background with Gd doping, where we assume a 90\% neutron tagging efficiency \cite{GdDoping}, therefore the background rate is reduced by a factor of ${\sim}3$. The dominant backgrounds are reduced, such as those from atmospheric muon neutrino charged-current interactions producing ``invisible'' muon decays. However, not all backgrounds are reduced, as neutrons can be produced from spallations as well as atmospheric neutrino neutral current quasielastic events. 

\begin{figure}[t]
    \centering
    \includegraphics[width=8.6cm]{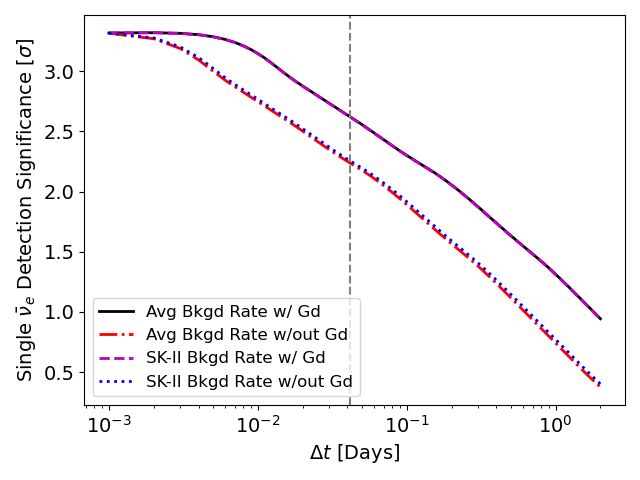}
        \caption{Plot of detection significance for a single neutrino event as a function of search window size assuming some background rate. The probability of detecting a neutrino due to backgrounds is found using the CDF of the Poisson distribution assuming that 0 events are detected. The vertical dashed line corresponds to a $\Delta t$ of one hour.} 
    \label{Detection Significance}
\end{figure}

We then define the significance as the $\sigma$-level corresponding to the probability of detecting one event due to backgrounds for a given search window size. This significance is shown in Fig.~\ref{Detection Significance}. As seen in the figure, SK-II versus averaged SK backgrounds makes little difference, while the addition of Gd has a large effect on the detection significance due to the background reduction it allows. It is important to note that these background rates for HK as they assume it will be similar to previous SK runs. As seen in Fig.~\ref{Detection Significance}, even with Gd a $5\sigma$ detection for a single neutrino is not possible with these background rates. Also, in Fig.~\ref{Detection Significance}, we can see how large an effect the uncertainty in time of core collapse has with how significant a detection can be. If there is no focus put on to increasing the precision of $\Delta t$, then for a single neutrino we cannot obtain confident detections with large search windows. 

For this close CCSNe approach, surveys like DLT40 should be able to get uncertainties in the time of core collapse to under one day. If HK has Gd doping and sees one neutrino from a CCSNe within 1 Mpc, that is roughly a $1.5{-}3$$\sigma$ detection, dependent upon the exact value of $\Delta t$. However, the core-collapse rate is so small that an event like this only happens once every few decades.

\subsection{CCSN neutrino detection rate}\label{LSST Combined Detection Rate Results}

We extend the CCSN neutrino search in the previous section by incorporating CCSN occurrence rates in estimating detection prospects. A simple extrapolation is to incorporate more CCSNe over longer periods of time. However, eventually a neutrino search will enter the uncorrelated DSNB regime. For example, assuming that the time of core collapse can be estimated to within 1 day, i.e., comparable to many survey cadence, we can at most search 365 CCSNe per year while remaining outside of the DSNB (continuous) regime. With more precise estimates of the core-collapse time, e.g., made possible by modeling the SN early light curve, one can consider more CCSNe. This is also a conservative estimate in terms of signal-to-noise as we assume that CCSNe happen at equal time intervals, when in reality their time intervals follow Poisson statistics. This means that the actual time to observe 365 CCSNe can be shorter as we can observe multiple CCSNe within a single time window, therefore the corresponding signal-to-noise can be enhanced. We quantify a combined detection rate as the summed total of neutrino events as:

\begin{equation}\label{Summed Neutrino Events}
    N_\mathrm{tot}=\sum^j_{i=1} N_{\bar{\nu}_e, i},
\end{equation}

\noindent where $j$ is the total number of CCSNe we are trying to study and $N_{\bar{\nu}_e}$ is the neutrino yield of each individual CCSNe given by Eq.~(\ref{NeutrinoYield}). As we detect an integer number of neutrinos from each CCSNe, this combined detection rate represents whether or not we can expect at least one neutrino event from a set of CCSNe. 

Using Eq.~(\ref{Summed Neutrino Events}) and the parameter values mentioned in Sec.~\ref{Core-Collapse Theory}, we calculate the event yields for the closest 365 CCSNe observable by ASAS-SN and LSST, retrieving 0.74 and 0.48 IBD events, respectively. ASAS-SN has more detected IBD events as it has a larger sky coverage than LSST, so it sees more CCSNe occurring at closer distances. However, since the forecasts are still ${<}1$, we consider studying more than 365 CCSNe. If we look at all possible observed CCSNe out to $z=0.075$ ($d\,{\sim}\,300$ Mpc), then ASAS-SN's CCSNe produce 1.05 IBD events and LSST's produce 1.42 IBD events. In this case LSST sees a larger IBD event yield as it has a large detection efficiency at these distances, whereas ASAS-SN is quite inefficient. However, to remain outside of the continuous DSNB regime, we must pinpoint the time of core collapse to smaller than one day. This can be done by modeling the early light curve of the SN. For example, treating it as a two-phase system with initial blackbody radiation from shock breakout followed by an expansion of a luminous shell, Ref.~\cite{ExplosionTime} used real SN light curve data (SN1987A, SN2006aj, and SN2008D) and determined the explosion time within an accuracy of better than a few hours. If we optimistically assume LSST will be capable of lowering the core-collapse uncertainty time to approximately an hour with such techniques, we can consider the closest ${\sim}9000$ CCSNe, increasing IBD event yield to ${>}1$ events. 

\begin{figure}[t]
    \includegraphics[width=8.6cm]{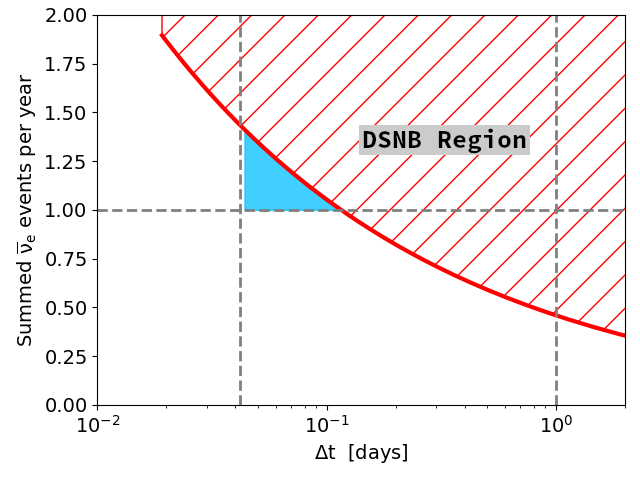} 
    \caption{Plot of LSST's achievable range of uncertainty of core-collapse time compared to summed $\bar{\nu}_e$ event numbers. The red hatched region represents the DSNB region where too many SNe are considered such that the neutrino signal is ``always on''. The blue shaded region is the ``region of interest'' where LSST can see enough CCSNe such that at least one neutrino is detected. There is an abrupt stop on the left for the red line which corresponds to all of the observed CCSNe our model predicts LSST can see in one year. 
    }
    \label{LSST Money Plot}
\end{figure}

With the above prescription, we identify a ``region of interest'' where LSST can observe enough CCSNe such that the combined detection rate is ${>}1$ neutrino. We show this region in Fig.~\ref{LSST Money Plot} as the shaded blue patch. The red hatched region is the continuous (DSNB) region where one would be studying more CCSNe than are allowed for a certain precision in time of core collapse ($\Delta t$). The horizontal dashed line corresponds to one IBD event, and the two vertical dashed lines correspond to $\Delta t$ values of one hour (left) and one day (right). We see that there is a reasonably sizeable area of the parameter space that allows for ${>}1$ detected events.  

Using the average background rate from SK and the neutron tagging efficiency brought by Gd doping, we can calculate the background of our theoretical CCSNe neutrino signal. The largest the background can be without our signal being in the DSNB regime assumes that we are considering the maximal amount of CCSNe for a given uncertainty in time of core collapse. The estimated background for this ``maximal'' signal region comes out to ${\sim} 77$ events per year assuming a Gd doped two tank HK with the same background rate as SK-II. This is much larger than our expected signal of ${\sim} 1$ CCSNe neutrino event. Even with perfect modeling, Poisson fluctuations will be $\sqrt{77} \sim 8.8 > $ signal rate. Therefore, this  coincidence search should be performed over several years of data so that only the closest CCSNe can be used, resulting in a better signal-to-noise. 

As an example, we estimate event yields corresponding to up to ten years of data taking with LSST and HK while studying the closest $N$ CCSNe. Extending to longer observation periods should help as the average distance of the closest $N$ CCSNe should decrease over longer times, thus giving a larger combined detection rate. For the backgrounds, we assume that the uncertainty in time of core collapse is one hour, thus making each search window an hour long. Assuming that HK will have a background similar to SK-II and that it will have Gd, this gives a detection significance of 2.62$\sigma$ for a single neutrino event. Therefore, we must make sure to look $N<113$ CCSNe, as that is the false alarm rate (FAR) for a false positive neutrino detection. With this, we chose to test different observation durations with the same number of CCSNe studied. The number of CCSNe we study has an upper limit given by the FAR and a lower limit which is somewhat arbitrary and chosen to be large enough such that we are not cutting out too many intermediate distance CCSNe. With those constraints in mind with our model, we chose to study the closest $N=100$ CCSNe. The resulting estimates for differing observation periods are shown in Table \ref{Events Table}.

\begin{table*}[t]
    \caption{\label{Events Table}Estimated summed $\bar{\nu}_e$ events for multiple years of data taking with LSST and HK. The closest 100 CCSNe are studied for neutrino event estimates. Also shown is the distance of the closest CCSNe. We do not study more CCSNe than this as we assume a one hour search window, giving ${\sim}2.62\sigma$ confidence (SK-II background with Gd doping), therefore we have a resulting false positive detection of one neutrino if we study more than ${\sim}113$ CCSNe.}
    \begin{ruledtabular}
    \begin{tabular}{ddddd}
    \multicolumn{1}{c}{Observation} & \multicolumn{1}{c}{Number of CCSNe} & \multicolumn{1}{c}{Closest CCSN} & \multicolumn{1}{c}{Average CCSN} & \multicolumn{1}{c}{Summed $\bar{\nu}_e$} \\
    \multicolumn{1}{c}{Years} & \multicolumn{1}{c}{Considered} & \multicolumn{1}{c}{Distance (Mpc)} & \multicolumn{1}{c}{Distance (Mpc)} & \multicolumn{1}{c}{Events}\\ \hline
    3 & 100 & 12.85 & 39.60 & 0.56 \\
    5 & 100 & 12.85 & 33.62 & 0.76 \\
    8 & 100 & 12.85 & 29.85 & 1.00 \\
    10 & 100 & 4.28 & 26.70 & 1.44 \\ 
    \end{tabular}
    \end{ruledtabular}
\end{table*}

We can see from Table \ref{Events Table} that the average distance for CCSNe decreases with increasing years spent observing. This is what causes the combined detection rate (Summed $\bar{\nu}_e$ events column) to increase when the same amount of CCSNe are considered with increasing years observed. Also, if enough time is spent observing, then closer CCSNe will occur, such as the closest distance only decreasing when ten years of observations are done. Once eight years of data is taken, the combined detection rate is ${\sim}1$ event when studying 100 CCSNe. So, we can expect that once eight years of data is taken, we can begin to expect neutrino detections from the intermediate distance regime. In terms of the number of background events, assuming the same rate as SK-II and a search window of one hour, the resultant background of a single window is 0.0066 events with Poisson fluctuations of 0.081. As previously mentioned, the significance of detecting one neutrino with this background is ${\sim}2.62\sigma$.

\section{Summary and Discussion}\label{summary discussion}

In this work, we have explored an approach to detect CCSN neutrinos in the intermediate distance regime between Galactic CCSN neutrinos and the DSNB. The approach relies on using a timing coincidence between the neutrino signal and the optical signal of the CCSN explosion. We use a combined detection rate to estimate how long observations need to be carried out for a measurable amount of neutrinos to be detected. For very long observations, a close enough ($D\leq1$ Mpc for our modeling) CCSNe will occur such that it should produce $\gtrsim6$ measurable neutrinos at a 2-tank Gd-doped HK configuration. A CCSN like this is a primary target for DLT40. For shorter observation periods, but still many years long, many CCSNe will occur at intermediate distances such that we can expect at least one neutrino to have been detected from a CCSNe within the observation period. For this, large coverage CCSNe surveys like ASAS-SN and LSST will provide the optical counterparts. 

For our forecasts, we modeled the predicted number of CCSNe ASAS-SN and LSST should be able to detect, then applied an analytic model of neutrino emission to each of these CCSNe to retrieve an integrated $\bar{\nu}_e$ flux at Earth over one year. Using the neutrino flux, we next calculated the number of detected IBD events. We then estimated the background rate by extracting the average background rate from SK's search for the DSNB \cite{SKbkgd}. Finally, we defined a search area in which future optical surveys and neutrino detectors can work in tandem to achieve a measurable integrated neutrino event flux from visible CCSNe. Our final results find that with waiting for $8{-}10$ years, a nearby CCSNe can produce a measurable number neutrino events at HK with our timing coincidence search method. Our analysis shows that for confident detections, small uncertainties in the time of core collapse are needed, and Gd doping greatly increases the significance of detections. The time frame for neutrino detections from this intermediate distance regime modeling does lie within the planned lifetimes of LSST and HK. 

There are various sources of uncertainty which can quantitatively upgrade our results in the future. The main theoretical uncertainties are in the CSFR (and hence CSNR) \cite{Horiuchi_2011, Horiuchi:2013bc} and the strength of dust extinction \cite{DustObscuration}, as well as the neutrino emission itself. In particular, the true CCSN rate in the local ${<}10$ Mpc is likely larger than the CSFR extrapolation we have used \cite{Ando_2005, Kistler_2011, Horiuchi:2013bc}, meaning our estimates are conservative. There are also unknowns with the experimental setup, for example the background rate, as HK will use different PMTs and photocoverage compared to SK. We also extract our background estimates from studies which do not go to as low an energy as we model, so further studies would need to be done to characterize the backgrounds between $11{-}15$ MeV with a focus on spallation as a few of them may be irreducible backgrounds even with Gd doping. Work has been done looking at using machine learning to help reduce backgrounds in this energy range, such as Ref.~\cite{Maksimovic_2021} with NC events. In our current modeling, a $5\sigma$ detection with a single neutrino event is not possible with our estimated background rate even with Gd doping. It is currently unclear whether HK will have a second tank, and whether the HK tank(s) will eventually have Gd doping \cite{hyperkamiokande}. Background rates will also impact the usable fiducial volume. We have assumed a conservative fiducial volume comparable to DSNB searches, but in a time-coincident neutrino search like the one explored, more of the HK volume may be usable. As the Vera C. Rubin observatory is also not yet finished, LSST has some inherent uncertainties as well \cite{LSST}. It is imperative that LSST focuses on the cadence and band availability to allow the uncertainty in time of core collapse to be narrowed down, as that directly determines how many CCSNe can be studied and how statistically significant a neutrino event at HK can be. 

There are possible future directions tied to expanding the search into other multi-messenger signals, such as gravitational waves (GWs) \cite{Yokozawa_2015, Vartanyan_2020, Mukhopadhyay_2021, Gill_2022, Powell_2022}, as they are not affected by dust or other visibility conditions. Planned upcoming GW experiments may be sensitive to the GWs produced during core collapse, especially if the collapse is asymmetric. These signal searches are also heavily reliant on the uncertainty in time of core collapse, as that constrains the GW search window. 

Offline searches for CCSNe neutrinos have already been performed at SK for distant CCSNe \cite{Ikeda_2007, SKIV_Search} and for DSNB signals \cite{SKbkgd, SKIV_DSNB}. The distant CCSNe studies rely on searching for event clusters, a collection of neutrino detections, within a time window characteristic of CCSNe, around 10 s. The search for DSNB signals at SK looks for unaccounted background signals that cannot be attributed to other processes. So far these searches have come up with null results. It is therefore warranted that offline searches of HK data will be done, and our suggested timing coincidence search would help with the signal-to-noise.

Collecting more CCSNe neutrinos is extremely important, as they can be used to test current theories of neutrino physics and processes of core collapse and progenitor dependence. Our only current repository of CCSN neutrinos all come from SN1987A, of which we have some two dozen events. With such a sample size, even the addition of a few more events from a few more CCSNe is meaningful. Hopefully within the coming decades, future optical surveys and neutrino detectors can probe the search area defined in this work, allowing the collection of more CCSNe neutrino events. 

\section*{Acknowledgments}

The authors would like to thank Chris Ashall and James DerKacy for very helpful discussions and introducing us to DLT40. S.~Heston is supported by NSF Grant No.~PHY-1914409. E.~Kehoe is supported by the National Science Foundation REU Grant No.~1757087. Y.~Suwa is supported by JSPS KAKENHI Grants No.~18H05437, No.~20H00174, No.~20H01904, No.~22H04571. S.~Horiuchi is supported by the U.S. Department of Energy Office of Science under Award No.~DE-SC0020262, NSF Grants No.~AST-1908960, No.~PHY-1914409 and No.~PHY-2209420, and JSPS KAKENHI Grant No.~22K03630.

\bibliography{research.bib}

\end{document}